\documentclass[conference]{IEEEtran}
\IEEEoverridecommandlockouts
\usepackage{cite}
\usepackage{amsmath,amssymb,amsfonts}
\usepackage{algorithmic}
\usepackage{graphicx}
\usepackage{textcomp}
\usepackage{xcolor}
\usepackage{balance}
\def\BibTeX{{\rm B\kern-.05em{\sc i\kern-.025em b}\kern-.08em
    T\kern-.1667em\lower.7ex\hbox{E}\kern-.125emX}}
\begin{document}

\title{Constructing Political Coordinates: Aggregating Over the Opposition for Diverse News Recommendation}

\makeatletter
\newcommand{\linebreakand}{%
  \end{@IEEEauthorhalign}
  \hfill\mbox{}\par
  \mbox{}\hfill\begin{@IEEEauthorhalign}
}
\makeatother

\author{\IEEEauthorblockN{Eamon Earl\IEEEauthorrefmark{2}\IEEEauthorrefmark{3}, 
Chen Ding\IEEEauthorrefmark{2}, 
Richard Valenzano\IEEEauthorrefmark{2}\IEEEauthorrefmark{3}, 
and Drai Paulen-Patterson\IEEEauthorrefmark{2}\IEEEauthorrefmark{3} }
\IEEEauthorblockA{
\texttt{\{eamon.earl,cding,rick.valenzano,dpaulen\}@torontomu.ca}\\
\IEEEauthorrefmark{2}\textit{Department of Computer Science} \\
\textit{Toronto Metropolitan University}, Toronto, Canada \\}
\IEEEauthorblockA{\IEEEauthorrefmark{3}\textit{Vector Institute}, Toronto, Canada\\}
}


\maketitle

\begin{abstract}
In the past two decades, open access to news and information has increased rapidly, empowering educated political growth within democratic societies. News recommender systems (NRSs) have shown to be useful in this process, minimizing political disengagement and information overload by providing individuals with articles on topics that matter to them. Unfortunately, NRSs often conflate underlying user interest with the partisan bias of the articles in their reading history and with the most popular biases present in the coverage of their favored topics. Over extended interaction, this can result in the formation of filter bubbles and the polarization of user partisanship. 
In this paper, we propose a novel embedding space called \textit{ Constructed Political Coordinates (CPC)}, which models the political partisanship of users over a given topic-space, relative to a larger sample population. 
We apply a simple collaborative filtering (CF) framework using \textit{CPC}-based correlation to recommend articles sourced from \textit{oppositional} users, who have different biases from the user in question. 
We compare against classical CF methods and find that \textit{CPC}-based methods promote pointed bias diversity and better match the true political tolerance of users, while classical methods implicitly exploit biases to maximize interaction.
\end{abstract}


\section{Introduction}

Recommender system (RS) utility has two main value measurements: users seeing content that they engage positively with, and the content providers maximizing engagement with their content or platform. While the two are evidently correlated (i.e. a user who is not properly catered to will likely cease to use the platform), the latter provides motivation for recommendation algorithms to shift a user's preferences to make them easier to cater to, resulting in higher expectations of long-term engagement \cite{induced_pref}.

Previous research \cite{poli_type} on the relationship between recommender systems and American political typology suggests that users with more extreme political preferences exhibit higher engagement metrics with their recommended news. Additionally, it was found that their engagement can be maximized by recommending articles among which a dominant percentage express a singular partisan bias. This establishes an implicit incentive for a News Recommender System (NRS) to shift user preferences toward political extremes through selection bias, particularly in long-term value systems or those leveraging popularity \cite{induced_pref}. This phenomenon results in the formation of filter bubbles, where users are eventually shown only perspectives in their news which comply with their pre-existing opinions, and users with heterogeneous partisanship over distinct topics have their political ideology homogenized over time. 

These effects are even more pernicious due to the algorithmic confounding inherent in recommendation: each iteratively deployed RS is trained on user histories that are conditioned by the selection bias of some previously deployed RS. In this case, not only are user preferences polarized and homogenized, but the models themselves are being trained on confounded data that incentivize this recommendation pattern, despite studies showing that homogenization effects lead to decreased user utility \cite{algo_confounding}.
The loss of political diversity in NRSs is ultimately a loss for civil discourse at large. In the context of social networks, there is a causal relationship between exposure to conflicting points of view on prominent issues and a general increase in political tolerance \cite{cross-cutting}. 

While traditional NRSs are responsible for the propagation of polarization in their user bases, the political bias lies in the content itself. Most Americans believe that news organizations often favor one side of the political spectrum and that organizations that favor the opposing side are less accurate in reporting the news \cite{pew_news}. Media bias can take many forms: content in favor of the incumbent government, in favor of a particular political party, ideologically liberal or conservative, in favor of specific industries, or in favor of audiences that are more valuable to advertisers \cite{media-bias}. It follows that individual articles in their perspective, language, and headlines often perpetuate this bias and leverage it to increase engagement from their desired audience. We focus on media bias towards ideologically liberal or conservative stances.

Current bias mitigation strategies often try to remove ideological bias from the representational space used by a RS \cite{cb_attention, disentangling}. In contrast, we explicitly focus on the bias present in articles to then infer the biases of the users who read them.
We argue that diverse news recommendation requires systems to directly model a user's underlying biases across distinct topics in order to provide the opportunity to interact with those opposing. Topic-specificity is critical: user partisanship must not be treated as monolithic. We define \textit{pointed diversity} as recommending articles that are diverse in their biases from articles of the same topic in a user's reading history.

In this paper, we leverage a bias disentangling model to construct a rich political embedding space, \textit{Constructed Political Coordinates (CPC)}, which places users according to their relative biases across prominent news topics\footnote{Code is available at: https://github.com/M4nd0C4lrissian/CPC}. We use this custom coordinate space to generate recommendations which naturally target political blind spots over distinct topics, providing articles with diverse perspectives from a user's reading history.

Our contributions can be summarized as follows:
\begin{enumerate}
    \item  We construct a novel political embedding space, \textit{Constructed Political Coordinates (CPC)}, which captures the political alignment of users across current topics.
    \item We propose \textit{Furthest Neighborhood in Political Coordinates} (FNPC), a method leveraging the informativity of \textit{CPC}-space via collaborative filtering to sample articles from oppositional users. 
    \item We provide in-depth analysis and display that \textit{FNPC} promotes \textit{pointed diversity} across distinct political typologies.
\end{enumerate}

\section{Related Work} \label{background}

The capability of NRSs to influence the broad political landscape beyond online spaces has been identified in numerous studies \cite{poli_type, effects_on_rec, stance_effects, div_pol}.
Interdisciplinary collaboration has called for the reconsideration of established definitions of news diversity: news outlets publishing diverse perspectives is no longer sufficient, the recommendation algorithm itself must also \textit{maintain} diversity \cite{diversity_in_news_rec}. Unfortunately, diversity of opinion within healthy democratic societies does not have a singular interpretation \cite{democratic_div} and is difficult to measure within online spaces \cite{measure_bubbles, radio}. 

Item diversity is generally acknowledged as an important metric to increase user satisfaction.
Early approaches to increase recommendation diversity applied combinations of nearest-neighbor and furthest-neighbor collaborative filtering algorithms \cite{fn, graph_conv}.
Diversity in news recommendation is unique: it has been established that not all users respond in the same way to diverse viewpoints. Users who are open to diverse viewpoints may receive less social reward in online spaces \cite{deliberate_exposure}. Some users may in fact become further polarized in the presence of diversity \cite{div_pol}. It was also found that avid news readers are more susceptible to being polarized by sentiments in their news content \cite{stance_effects}. Treuillier et al. \cite{multi_fac, div_not_enough} proposed modeling user classes via their polarization behaviors and topic stances in order to personalize the degree of diversity introduced.
Alam et al. \cite{stance_pref} showed that RSs can favor recommending specific stances on a given topic, often correlated with the pre-existing majority stance among users. Additionally, opposing stances over a single topic have been found to induce different effects: stances \textit{against} an idea were more often found in filter bubbles than stances in support of it \cite{viewpoints}. 

Anwar et al. \cite{bubble_or_homog} distinguished inter-user diversity (two users having distinct consumption patterns) and intra-user diversity (a single user consuming diverse content). They showed how the two can be traded off by recommending a curated (non-personalized), diverse article set which maximized intra-user diversity while minimizing inter-user diversity. Our work differs in that we aim to achieve \textit{pointed diversity}, which requires targeting articles that are diverse in their bias from articles of the same topic in a user's immediate history. Thus, the diversity introduced to each user is dependent on their prior biases on different topics. This can be seen as a manner of increasing intra-user diversity in a personalized manner, and thus without homogenizing article recommendations.

Recent efforts have been made to model heterogeneous user interests in a manner that is devoid of article biases. Liu et al. \cite{transformer_debias} proposed a transformer network to detect and neutralize polarized language from article content. 
Shivaram et al. \cite{cb_attention} trained an attention network to attend to topic-predictive terms and to ignore biased terms that appeared across many topics. 
Wang et al. \cite{disentangling} applied adversarial autoencoders to isolate the polarity-free information from news articles. 
What is not considered by these approaches is that topics are often covered disproportionately by left-leaning or right-leaning outlets \cite{polarization_news}, where bias can be reflected in the presence, lack, or sequence of specific information \cite{propaganda} regardless of the language used. Different topics can also have different polarization effects \cite{polarity_topics}. Removing awareness of bias-predictive information from a recommendation algorithm could then promote arbitrarily singular biases for such topics.

Inspired by the work of Treuillier et al. \cite{div_not_enough}, we create informative models of user bias on different topics. We then define a coordinate space where directional dissimilarity is correlated with difference in perspective. Using this space, we can sample articles from \textit{oppositional} users to recommend content that confronts the biases of the user we are serving in a personalized manner.

\section{Political User-Space Simulation} \label{user} 

In this section, we discuss a user simulation framework and dataset \cite{poli_type} that allow us to analyze how our \textit{CPC}-space models distinct ideological classes, referred to as \textit{typologies}. 


Liu et al. \cite{poli_type} proposed a framework and dataset for generating user profiles with political biases proportional to those of the American public. 
We apply this framework to show that relative to the greater \textit{body politic}, our \textit{CPC} embedding space captures meaningful information regarding a user's political biases across distinct topics. 

The authors generated a custom dataset composed of articles from 41 American news outlets\footnote{Dataset is available at: https://github.com/IIT-ML/WWW21-FilterBubble}. Each article was associated with a political stance score among \{-2, -1, 0, 1, 2\} based on the news outlet that produced it, ranging from very liberal (-2) to very conservative (2). The stances were sourced from \textit{www.allsides.com}, and as such are influenced by both mass public opinion and expert assessment \cite{allsides}. An even distribution of the 5 political stance scores was maintained among the 40,000 articles in the dataset.
The authors identified a set of 14 topics (as shown in Table \ref{tbl:dataset_topics}) that encapsulated the relevant cross-cutting issues in 2020 American politics. A combination of expert annotators and NLP methods were used to classify articles into topics, but only human perspectives influenced the bias scores.
 For each article, a binary article matrix $A$ was constructed. Each matrix $A$ is of dimension 14x5, where $a_{ij} = 1$ if the article covers topic $i$ with a partisan bias of index $j \in$ \{-2, -1, 0, 1, 2\} + 2 (the index of each stance score is shifted to be positive). 

\begin{table}
\centering
\caption{14 topics in the dataset and their distributions}        \label{tbl:dataset_topics}
\begin{tabular}{|l|c|l|c|}
\hline
 Topics & \% Articles & Topics & \% Articles \\
 \hline
abortion & 2.8 & environment & 3.5 \\
guns & 3.7 & health care & 10.9 \\
immigration & 9.8 & LGBTQ & 2.5 \\
racism & 8.2 & taxes & 5.7 \\
technology & 2.7 & trade & 4.8 \\
Trump impeachment & 11.8 & US military & 15.3 \\
US 2020 election & 14.5 & welfare & 3.8 \\
 \hline
\end{tabular}
\end{table}

This dataset is useful for two reasons. First, the perceived bias of the news is sourced directly from the perspective of American citizens. 
Individual human biases are the foundation for effective media bias, and as such, we are naturally tuned to identify the symbols that are relevant in predicting it, especially when explicitly prompted \cite{propaganda, hostile_media}.  While deep learning methods have achieved some success in this task, they still suffer from sampling bias and a lack of interpretability in their predictions \cite{media_bias_survey}.
Second, the dataset has 14 discrete, politically relevant topics within which all article content can be classified. This allows us to  analyze how \textit{CPC} enables \textit{pointed diversity} over these distinct topics.  

The authors leveraged survey data gathered by a Pew research study, which proportioned the general American public into 9 political typologies: \textit{bystanders} (5\%), \textit{core conservatives} (15\%), \textit{country first conservatives} (6\%), \textit{devout and diverse} (8\%), \textit{disaffected democrats} (12\%), \textit{market skeptic republicans} (11\%), \textit{new era enterprisers} (10\%), \textit{opportunity democrats} (13\%), and \textit{solid liberals} (20\%) \cite{pew_typology}. The Pew research study recorded statistics of agreement among each political typology with regards to opinions on cross-cutting topics. To create the user generation scheme from this survey data, Liu et al. \cite{poli_type} first isolated the survey questions corresponding to the 14 topics in the dataset. 
They calculated the dominant bias of a given typology for a given topic by associating each bias score in \{-2, -1, 0, 1, 2\} with a percentile (0-20\%, 21-40\%, etc.) of agreement on questions regarding that topic (i.e. if 92\% of solid liberals agree that abortion should be legal in all/most cases, then their dominant bias score will be -2 on abortion). 

This data is then used by Liu et al. \cite{poli_type} to simulate individual user biases over all topics.
A user is modeled with a \textit{user bias matrix}, which is a 14x5 matrix $UB$ of bias-topic values $b_{i,j} \in [0, 1]$. Each $b_{i,j}$ indicates the user's probability of interaction with an article due to solely its topic $i \in [0, 13]$ and perceived bias $j \in$ \{-2, -1, 0, 1, 2\} + 2. 
For example, a solid liberal user may have the following row in their $UB$ matrix for the topic of abortion: $[0.94, 0.80, 0.65, 0.21, 0.10]$, which indicates that they have a 94\% chance to interact with a very liberal article on the topic of abortion, and only a 10\% chance to interact with a very conservative one. 

To generate the \textit{bias matrix} entry $b_{t, d}$ for the dominant bias index $d \in$ \{-2, -1, 0, 1, 2\} + 2 over a given topic $t$, they sample from a Beta distribution $Beta(p_{t}, 1)$, where $p_{t}$ is the average percent agreement among the political typology of the user to be generated (i.e. for $t = 0\rightarrow \verb|abortion|$ and a solid liberal user where $d = 0$, $b_{0, 0} \sim Beta(0.92, 1)$). To account for variation among members of a given political typology, the interaction probability of the dominant bias score $b_{t, d}$ is decayed to neighboring bias scores as a function of the standard deviation in survey responses. Thus in the previous example, the value of 0.94 was sampled directly from $Beta(0.92, 1)$, which was incrementally decayed to derive 0.80, 0.65, 0.21 and 0.10. This process is repeated for each topic to generate the \textit{user bias matrix}, $UB$, of a single user.

Despite being sourced from survey data taken from a sample of the American populace, there is no explicit guarantee that any simulation framework is exactly reflective of reality. The key element of these \textit{user bias matrices} that we make use of is that they capture heterogeneous partisanship over distinct topics and thus can be treated as coherent representations of the broad ideological identities that make up American society. 
Additionally, these interest models are not directly confounded by the recommendation patterns of any prior NRS, and thus arguably capture a more realistic representation of the heterogeneity of partisanship among the general public when not influenced by online spaces.

For a user $u$ with bias matrix $UB$, the probability $p_{uv}$ of interaction with a given article $v$ is the dot-product of the flattened $UB$ matrix and binary article matrix $A$. In other words, it is the average interaction probability $\bar{b}_{\cdot, s}$ over all topics covered in $v$, where $s$ is the bias score of the article.
To simulate user histories, Liu et al. \cite{poli_type} uniformly draw sample articles $v$ from the item pool, and log an interaction with probability $p_{uv}$.

Via this approach, articles with similar perspectives to a user's maximal bias preferences will be more likely to be interacted with. Additionally, interactions will include a greater proportion of the most popular topics, as is expected in news recommendation environments. Due to these two phenomena, Liu et al. \cite{poli_type} show that classical RS methods homogenize user interest towards their particular partisanship on the most popular topics, ignoring their biases over niche topics.


\section{Methodology} \label{method}
In this section, we first introduce the bias disentangling module. Then, we outline the formulation of our \textit{Constructed Political Coordinates} and show how this novel embedding space can be leveraged to target \textit{pointedly diverse} recommendations in a simple GCF model (furthest-neighbor method). Finally, we provide details on the training process for the bias disentangler and GCF models.

\subsection{Bias Disentangling} \label{bias_disentangling}

Previous work in news recommendation \cite{cb_attention} has shown that removing the influence of bias in content representations can improve recommendation diversity. One such method \cite{disentangling} involves applying an adversarial autoencoder model to produce two content embeddings, one that contains bias-predictive information and one that does not. The previously proposed method then discards the bias-predictive embedding, and provides only the \textit{depolarized embedding} to the model for recommendation. As discussed in Section \ref{background}, this method does not explicitly target diverse biases on topics of interest and can still provide recommendations that are arbitrarily singular in their biases.

As opposed to discarding the polarized article representations, we explicitly leverage this bias-predictive information in order to generate our \textit{Constructed Political Coordinates}. The intuition is that by modeling user interests relative to the manner in which they are influenced by bias, we can recommend articles that specifically confront those biases. We believe that biases will inevitably be present in news articles and that providing a balanced measure of bias over a given topic is the best way to meaningfully address its influence. 

In Fig. \ref{fig:bias_dis}, we show a high-level representation of inference in the bias disentangling module. Firstly, a pre-trained BERT \cite{devlin-etal-2019-bert} model is applied to generate embeddings over article titles and descriptions, which are combined and compressed via an attention layer. From there, an encoder maps the input vector $c$ of dimension $M$ to the latent space of dimension $m$. Then, two different decoders are applied to the latent vector with the intention of disentangling the bias-predictive information contained in the input vector $c$. Each decoder produces a `half' embedding of size $M$ / 2, denoted $D_p$ and $D_f$. We subscript the decoder outputs with $p$ and $f$ to signify whether it is trained to contain \textit{polarized} (bias-predictive) or polarity-\textit{free} information, respectively. 

In order to enforce that bias-predictive information is encoded in $D_p$ and is absent in $D_f$, a classifier $C$ is used as a discriminator. Provided $D_p$ as input, the classifier learns to predict the true bias label $y \in \{-1, 0, 1\}$, corresponding to `left', `center', and `right'. Conversely, when $D_f$ is the input, the classifier is trained to fail to predict $y$. The classifier produces a probability distribution over labels:
$C(D) = \hat{\mathbf{y}} \in [0,1]^3$ where $\sum_i \hat{y}_i = 1$.
$C$ is trained using one-hot label vectors $\mathbf{y}$.
Thus, the training objective of the bias disentangler comprises three tasks learned in tandem: 
\begin{align} 
&L_{total} = L_{class} + L_{conf} + L_{recon}\\
&= - log (C(D_p)_j) -\frac{1}{log (C(D_f)_j)} + \frac{1}{2}(c - [D_p ; D_f])^2 \label{disentangler loss} 
\end{align}

The last term in Equation \ref{disentangler loss} is the MSE of the reconstruction of the input vector $c$, which is predicted via concatenating the embeddings $D_p$ and $D_f$. This term is necessary to maintain coherent representations of the input. The former two terms are  classification loss and confusion loss (cross-entropy and its reciprocal form), where we assume the one-hot encoding of our true label is 1 at index $j$.

\begin{figure}[htp]
    \centering
    \includegraphics[width=\columnwidth]
    {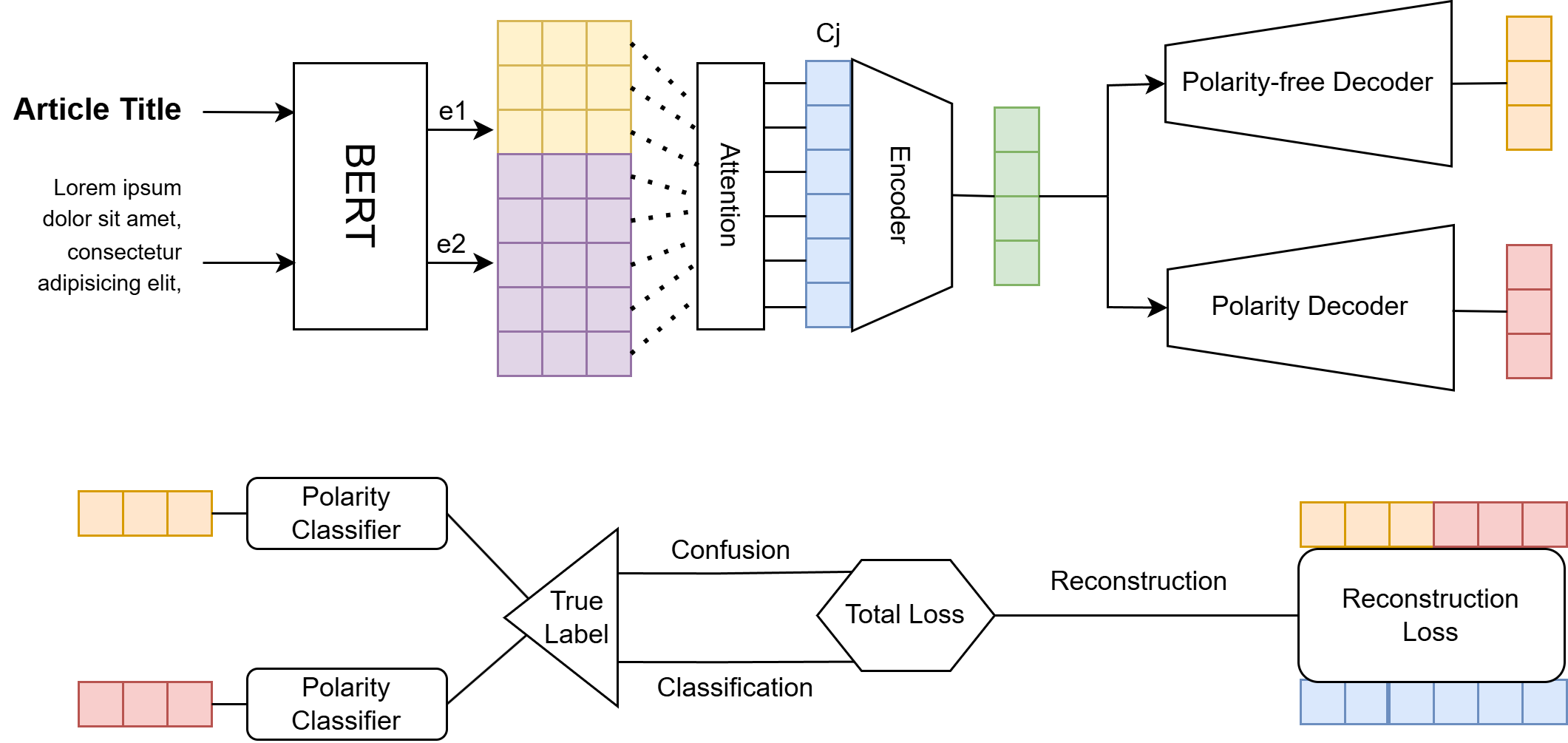}
    \caption{The Bias Disentangling Module}
    \label{fig:bias_dis}
\end{figure}

\subsection{Constructed Political Coordinates} \label{CPC}

For every user $u$, we can leverage their generated history $H_u$ to create a polarized user embedding $W(u; H_u)$ in the representational space of our polarized decoder.
\begin{align} 
W(u ; H_u) = \frac{ \sum_{v_i \in H_u} p_{uv_i} \cdot D_p(v_i) }{ \sum_{v_i \in H_u} p_{uv_i} }
    \label{polarized_emb}
\end{align}

The polarized user embedding is calculated as a combination of the bias-predictive information in their previously-read articles, weighted by the interaction probabilities $p$. This results in a user embedding of dimension $M$ / 2. We use the $p$ values to represent a measure of satisfaction that the user expresses with each logged article, thus we refer to them hereafter as \textit{ratings}. This weighting ensures that user bias is measured in greater proportion with the articles they most enjoyed.

As opposed to constructing our political coordinates in this high-dimensional space, we further compress our coordinate space using \textit{landmark users}. For the purposes of this study, we derived 9 \textit{landmark users} by calculating the average \textit{user bias matrix} among the users of each political typology respectively. We then followed the method outlined in Section \ref{user} to generate the histories of our \textit{landmark users} $l \in L$, and generated \textit{landmark embeddings} $W_l(l ; H_l)$  using Equation \ref{polarized_emb}.

\begin{sloppypar}
In order to generate our \textit{Constructed Political Coordinates} (CPC) representation of a given user $u$, we compute the user-landmark differences $W(u; H_u) - W(l; H_l)$ for each landmark user $l$. We then normalize each dimension $i < M / 2$ of these 9 vectors by the range $\underset{l \in L}{\max}[W(l)_i] - \underset{l' \in L}{\min}[W(l')_i]$. This normalization defines an axis-aligned bounding box (AABB) that contains all landmark embeddings along each of the $M$ / 2 dimensions. If any normalized component of a user-landmark difference vector exceeds the range $[-1, 1]$, it implies that the user differs more from a given landmark in that dimension than any pair of landmarks differ from each other.
\end{sloppypar}
 To compress these 9 high-dimensional difference vectors into a compact coordinate representation, we compute the L2 norm of each normalized vector. This produces a 9-dimensional vector which we refer to as a user's \textit{Constructed Political Coordinates}. \textit{CPC} captures the distance of a user from the most standard users of each typology.  
As the normalization factor reflects the full dimension-wise extent of the landmark set, users whose normalized difference vectors have all components within $[-1, 1]$ fall within the axis-aligned bounding box defined by the landmarks. Users with one or more components exceeding this range lie outside this region of observed landmark variation. 
As such, the L2 norms encoded in our \textit{CPC}-space are proportional to the centrality of a user's ideology: values closer to 0 mean that the user is better-contained within the ideological space described by our landmark users, while larger values indicate a user that is closer to the fringe of the greater \textit{body politic}.

\subsection{Furthest-Neighborhood Aggregation}
Once we construct our \textit{CPC}-space, we use it to calculate user correlation and create user-adjacency matrices. We then apply Graph Convolutional Filters (GCFs) to aggregate user interactions and generate recommendations.

In a GCF, predicting user ratings of an item $i$ with adjacency matrix $B$ means shifting logged ratings along the paths defined by $B$. We can perform $k$ shifts or `hops', equivalent to $\hat{x} = B^k x_i$ where $x_i$ is a vector containing all user ratings for item $i$. 
The GCF model introduces learnable parameters $h = [h_0, ..., h_k]^T$ to weigh information at each resolution according to:
\begin{align} \label{graph_convolution}
\hat{x} = (\sum_j^kh_jB^j)\cdot x_i
\end{align}

Essentially, we compute all of the $k$-step paths from all users, and propagate rating information back along those paths, weighted by the $h$ weights at each layer and by the weights of the edges between user nodes. 

We train GCFs to aggregate rating information and generate candidate articles. We compile user ratings into a user-item matrix $X$ of dimension $|U|$ x $|V|$, where $U$ is the set of all users and $V$ is the set of all items. For each user $u$, if they interacted with an item $v$, then $X_{u,v} = p_{uv}$, else $X_{u,v} = 0$.  

We create the correlation matrix CPC-Corr by calculating the Pearson correlation coefficient between our users in their \textit{CPC} representations. This means that users with similar patterns of distance to each \textit{landmark user} will be considered highly-correlated, irrespective of their magnitude of distance. In other words, highly-correlated users follow a similar directional profile: they deviate from our landmark users in similar ways. 
We construct two distinct adjacency matrices: the furthest-neighbor matrix $F$, and the nearest-neighbor matrix $N$. To do so, we retain at row $i$ of both matrices the correlation scores of the $n$ most negatively and most positively correlated users respectively, according to the CPC-Corr matrix. The remaining correlations are zeroed out. 

We construct a variant of the GCF where rating prediction uses both the $F$ and $N$ matrices as seen in Equation \ref{FF_CPC}. This means that our first hop targets our \textit{oppositional} users, and subsequent hops target the closest neighbors of this opposition group. Thus to predict the ratings of a given user, we aggregate ratings from the \textit{furthest-neighborhood} of that user.
\begin{align} \label{FF_CPC}
\hat{x} = h_1F \cdot (\sum_{j=2}^kh_jN^{j-1})\cdot x_i.
\end{align}

We refer to the resulting predictive model as the \textit{Furthest Neighborhood in Political Coordinates} (FNPC) method.

The intuition behind \textit{CPC} is that articles will inevitably be infused with bias; by modeling user interests relative to these biases we can recommend articles which specifically confront them. Thus, by targeting furthest-neighborhoods in \textit{CPC}-space, \textit{FNPC} can achieve \textit{pointed diversity}. 

We compare our method with a vanilla nearest neighbors (\textbf{NN}) baseline. In this method, the correlation matrix is calculated as the Pearson correlation coefficient over user ratings, which does not leverage \textit{CPC}-space. As such, its prediction is exactly equivalent to Equation \ref{graph_convolution}, where $B = N_X$, the nearest-neighbor correlation graph calculated over $X$.

\begin{table*}[h]
\centering
\caption{Top-5 recommendations from \textit{FNPC} for a user of class ``new era enterprisers'' \& the user's previous historical articles from the recommended topics. }        
\label{tbl:recs}
\begin{tabular}{l|c|c}
\hline
\textbf{History} & \textbf{Topic} & \textbf{Bias} \\
\hline
"Coronavirus Crisis Spurs Access To Online Treatment For Opioid Addiction" & healthcare & 0\\
"Elizabeth Warren misleads Americans on harm and costs of 'Medicare-for-all'" & healthcare & 1\\
"House chairmen demand explanation on Trump's 'illegal' withdrawal from Open Skies Treaty" & military & 0\\
"U.S. forces could still be at risk in Syria, defense secretary says" & military & 1\\
"Hillary Clinton Endorses Biden Amid Developments In Reade Allegations" & election & 2\\
\hline
\textbf{Top-5 Recommendations} & \textbf{Topic} & \textbf{Bias}\\
\hline
"5 ways to prevent another 100,000 coronavirus deaths in the US (and beyond)" & healthcare & -2\\
*"Opinion | Here's a playbook for defeating the NRA in the wake of the Saugus shooting" & *healthcare & -1\\
"U.S. envoy to Afghanistan announces 'pause' in Taliban talks after attack on Bagram air base" & military & -1\\
"U.S. Companies Get Slice of Iraq's Oil Pie" & military & -1\\
"How's Trump Doing with the Black Vote? A Lot Better Than You Might Think." & election & 2\\
\hline
*This article considers gun control from a public health perspective.
\end{tabular}
\end{table*}

\subsection{Implementation Details} \label{details}

We generate 1,000 users according to the generation scheme and typological distribution outlined in Section \ref{user}. 
We follow the method outlined in Section \ref{user} to log 20 interactions per user from a set of 4,000 articles over which the distribution of bias scores is even. We choose a relatively small number of interactions to reflect that many articles and entire topics very quickly become out-of-date. As such it is not reasonable to assume a long history over a single set of articles or topics.

Provided the dataset discussed in Section \ref{user}, we apply the base BERT model to preprocess the article titles (64 tokens) and descriptions (256 tokens) into token embeddings of dimension 768.
Our disentangler module was trained using a distinct set of 32,000 articles, with only 600,000 parameters adjusted during the main training process. Thus, the construction of our \textit{CPC} is fairly lightweight. 

For training the disentangler, text and title embeddings were first concatenated into a single matrix of dimension 320 x 768 and fed into an attention layer with 8 attention heads and a dropout of 0.1. The attention layer casts the concatenated inputs into a single embedding vector $c$ of dimension 256.
Our encoder is a dense network with one hidden layer employing both residual connections and LeakyReLU activation functions. The encoder takes in vector $c$ as input and produces a latent representation of dimension 128. 
The two decoders have the same architecture with reversed dimensions, except that their output dimensions are 256 / 2 = 128.
The \textit{polarity classifier} is also a dense network with a single hidden layer and a softmax activation function.
We applied the Adam optimizer with a learning rate of 0.001 and a batch size of 50 over 32 epochs, with each epoch covering a different 1,000 articles from our training set. 
We pre-trained only our attention layer and encoder on the first 3,000 articles in the training set to maximize the orthogonality between latent representations of articles with opposing bias scores. This incentivized latent representations to reflect bias-predictive information, and provided more informative initial loss topography. We froze our attention layer for the main training process. 

For each user, we retain half (10) of their interactions as holdout ratings that are not used for constructing our \textit{CPC}-space, calculating user correlation, or predicting ratings. A set of GCF weights $h$ are trained per user to minimize the MSE between the holdout ratings and those predicted by the GCF aggregation via Equation \ref{graph_convolution} or \ref{FF_CPC}. We recommend the 10 unread articles with the highest predicted ratings for each user. 
For both tested methods and each of our 1,000 users we retain 8 neighbors in the respective $F$ and $N$ matrices and train a 5-hop GCF over 40 update steps using Stochastic Gradient Descent. We consider this feasible due to the lightweight nature of singular GCFs.

In Table \ref{tbl:recs} we show an example of the top-5 recommendations from our model and all the articles with matching topics from the user's previous interaction history. This example exhibits the pointed nature of the topic-specific diversity introduced by our method.

\section{Empirical Evaluation} \label{pointed}

\begin{table*}

\caption{Click-through rate, Wasserstein distance, political tolerance, normalized entropy, and average bias.} \label{tbl:stats}
\centering
\begin{tabular}{|l|cc|cc|c|ccc|ccc|}
\hline
      \textbf{Political} & \multicolumn{2}{c|}{\textbf{CTR}}  & \multicolumn{2}{c|}{\textbf{WD}} & \multicolumn{1}{c|}{\textbf{PT}} & \multicolumn{3}{c|}{\textbf{NE}} & \multicolumn{3}{c|}{\textbf{AB}}\\
       \textbf{Typology} & NN & FNPC & NN & FNPC & & NN & FN-NN & FNPC & NN & FN-NN & FNPC\\
       \hline
       bystanders & 0.41 & 0.37 & 1.35 & 1.53 & 0.98 & 0.74 & 0.60 & 0.66 & -0.90 & -0.21 & -0.30\\
       \hline
       solid liberals & 0.74 & 0.40 & 0.81 & 1.68 & 0.84 & 0.36 & 0.60 & 0.71 & -1.56 & 0.05 & 0.05 \\
       \hline
       opportunity democrats & 0.66 & 0.51 & 0.97 & 1.52 & 0.93 & 0.42 & 0.61 & 0.70 & -1.43 & -0.05 & -0.46 \\
       \hline
       disaffected democrats & 0.60 & 0.48 & 1.03 & 1.50 & 0.94 & 0.46 & 0.60 & 0.70 & -1.20 & -0.03 & -0.43 \\
       \hline
       devout and diverse & 0.42 & 0.33 & 1.42 & 1.65 & 0.99 & 0.67 & 0.60 & 0.70 & 0.45 & -0.24 & -0.28 \\
       \hline
       new era enterprisers & 0.51 & 0.43 & 1.16 & 1.51 & 0.97 & 0.62 & 0.58 & 0.72 & 0.85 & -0.24 & -0.22  \\
       \hline
       market skeptic repubs & 0.63 & 0.41 & 1.06 & 1.61 & 0.97 & 0.59 & 0.58 & 0.68 & 0.98 & -0.21 & -0.30  \\
       \hline
       country 1st conservs & 0.69 & 0.36 & 0.93 & 1.96 & 0.88 & 0.50 & 0.56 & 0.71 & 1.33 & -0.32 & -0.52 \\
       \hline
       core conservatives & 0.77 & 0.37 & 0.82 & 1.83 & 0.87 & 0.45 & 0.58 & 0.68 & 1.39 & -0.35 & -0.38\\
       \hline
        \textbf{Avg.} & 0.64 & 0.41 & 1.00 $\pm$ 0.42 & 1.64 $\pm$ 0.65 & 0.92 & 0.50 & 0.59 & 0.70 & $|1.20|$ & $|0.17|$ & $|0.30|$ \\
        \hline
\end{tabular}

      \vspace{5px}
      \small
      We only report standard deviation of WD due to space constraints.
      The global average AB scores are taken over absolute values.
\end{table*}

\subsection{Metrics}

To assess the quality of \textit{pointed diversity}, we must  measure the difference in article bias between those recommended and those previously read. To do this, we consider the Wasserstein distance (WD) between the per-topic bias distributions induced by the reading history and the recommended articles respectively. This metric measures the minimum distance one would have to shift probability mass over a support in order to equate two different distributions. Given that we have a discrete support of \textit{relative} bias scores \{-2, -1, 0, 1, 2\}, this distance will be highly interpretable: in our case, WD is bounded between 0 and 4.

To calculate WD, we first have to produce the per-topic bias distributions.
For a given topic $t$ and user $u$ we create a count $C_{t, j}$ of the number of articles in their interaction history $H_u$ that cover $t$ with bias score $j \in $ \{-2, -1, 0, 1, 2\}.
This generates a histogram over bias scores, which we normalize to sum to 1. The result is a distribution, $U_t:$ \{-2, -1, 0, 1, 2\} $\rightarrow [0, 1]$, representing the posterior likelihood of a user interacting with articles of topic $t$ as a function of the article bias, conditioned by their history. In an identical fashion we calculate the per-topic bias histogram of the recommended articles, which we normalize to form the distribution $R_t$. The discrete 1-Wasserstein distance over a common support is then calculated according to Equation \ref{wass_1}, where $F_P$ is the cumulative mass function of arbitrary distribution $P$.
When reporting WD, we average this distance per user over solely the topics present in both the recommendations and prior interaction history. By comparing the distributions for each topic separately, we can see how well an RS manages to model heterogeneous interest. 
\begin{align} \label{wass_1}
    WD (U, R) = \sum_{b=-2} ^2 |F_u(b) - F_r(b)|
\end{align}

We note that a low WD for a RS that seeks maximum engagement (i.e. the vanilla NN) will indicate successful heterogeneous modeling. For \textit{FNPC}, a relatively higher score is indicative that \textit{pointed diversity} has been achieved.
We argue that the ideal WD that is indicative of \textit{pointed diversity} is between 1.5 and 2. A score of 1 or less implies that the recommendations are mostly on the same side of the political axis as the prior articles (i.e. if a prior distribution was concentrated on -2 and -1, a WD of 1 means the recommendations were concentrated on -1 and 0). In other words, it signals a general lack of bias diversity. An average WD notably greater than 2 means that recommendations are generally on the opposite side of the axis, but are also the most polarized articles from that side. Scores between 1.5 and 2 indicate that oppositely leaning articles are consistently being recommended, but they are not exclusively -2 or +2 in their bias. In this way, more extreme perspectives are shown as well as more moderate ones.

We also consider how RSs exploit the biases of users to maximize interaction. To do so, we consider the well-established metric click-through rate (CTR), a measurement of the proportion of recommendations that a user interacts with. This is a measure of the average $p_{uv}$ score of recommended articles. User histories are populated in greater proportion with articles whose topic-bias pairs induce a high probability of interaction. As such, the goal of \textit{FNPC} in this case is to recommend articles whose topic-bias pairs have low probabilities of interaction, and thus to receive low CTR scores.

As noted by Treuillier et al. \cite{multi_fac}, the normalized entropy (NE) of perspectives present in a recommendation distribution is an effective measure of polarization, where lower values indicate a more polarized set of articles. Equation \ref{rec_entropy} shows how we calculate NE according to the proportion $q_j$ of articles recommended to a user of bias score $j$. We normalize this value by $log \, 5$ to bound the result in [0,1]. 
\begin{align} \label{rec_entropy}
 NE = \frac{-\sum _{j=-2}^2 q_j \, log \,q_j}{log \, 5}
\end{align}

 This metric in addition to WD can help us validate whether \textit{pointed diversity} is being achieved in a manner that does not solely target the most polarized articles from \textit{opposition} users. 

Validation of our approach requires showing that the nature of the depolarization achieved is due to our correlation being calculated using \textit{CPC}-space, and not solely due to the manner in which ratings are predicted. Thus, we perform an isolated evaluation of the properties of the \textit{furthest-neighborhood}.

We again apply Equation \ref{FF_CPC} to target the \textit{furthest-neighborhood} of a user, using $F_X$ and $N_X$ matrices whose correlation was calculated directly over the user-item matrix $X$, and not using our \textit{CPC}-space. While this method should also target diverse biases, it will do so in a manner that caters primarily to the dominant typologies, and can be expected to provide highly polarized recommendations with lower entropy than \textit{CPC}. We refer to this method as \textbf{FN-NN}.

We are interested in comparing the entropy of recommendations to the prior entropy of users according to their \textit{user bias matrices} $UB$ discussed in Section \ref{user}. To do so, we sum up the interaction probabilities for each bias score over all 14 topics: $\sum_{i=0}^{13}UB_{ij} \, , \, j \in \{-2, -1, 0, 1, 2\} + 2$. We then normalize the resulting per-bias values to form a distribution that captures the proportion of articles of each bias that a user would interact with if recommendations were sampled from a uniform distribution over topic-bias pairs. We measure the normalized entropy of this distribution, and refer to the resulting value as the \textit{political tolerance} (PT) of a user.
We additionally measure the average bias score (AB) of recommendations per user, averaged over recommended topics. Higher magnitudes of AB indicate that an RS is disproportionately recommending articles from one side of the political spectrum.

\subsection{Analysis} \label{analysis}

For the purpose of analyzing the results, it is useful to first report some general characteristics about the 9 political typologies. While this will not be comprehensive, it will establish a framework of analysis. Firstly, \textit{solid liberals} and \textit{core conservatives} comprise the greatest majority of Americans and represent the core identity of the `left vs. right' paradigm. It is worth noting that \textit{solid liberals} are more consistently left-leaning than \textit{core conservatives} are right-leaning.
On the ideological right, \textit{country-first conservatives} differ from this core in that they are highly right-leaning on issues of immigration, racism and LGBTQIA. \textit{Market-skeptic republicans} differ from the core due to their left-leaning views on technology and taxes. \textit{New era enterprisers} are right-leaning on tech and taxes, and left-leaning on immigration.
On the ideological left, \textit{opportunity democrats} differ from the \textit{solid liberals} via their right-leaning views on technology and taxes. The \textit{disaffected democrats} mostly align with the leanings of \textit{solid liberals}, but are overall more central and less politically involved. 
\textit{Devout and diverse} Americans are difficult to place on either side, and particularly heterogeneous in their partisanship over topics. They are left-leaning on issues of racism, healthcare, and welfare, but right-leaning on issues related to US economy and military, as well as on issues of immigration and LGBTQIA.
The final typology is the \textit{bystanders}, who are politically disengaged, and generally read articles with more central bias. 
When reporting metrics averaged across typologies, we weigh the respective values by the proportion of that typology among our 1,000 users. 

Table \ref{tbl:stats} shows the results of all evaluation metrics for each political typology. If we analyze CTR scores for the classical NN approach, we see that the two highest scores belong to the \textit{solid liberals} (0.74) and \textit{core conservatives} (0.77). These typologies are the most singular in their beliefs, especially over the most popular topics. This means that their interaction histories are limited to the subset of articles whose biases align with their singular viewpoint. Thus they will be highly correlated over ratings to other users who interact with these singularly biased articles, who will very often be of their own typology. 
We see consequently that the average bias (AB) of the NN method is very typology-dependent and the magnitude is quite high, especially for the most polarized and dominant typologies.
As intended, our \textit{FNPC} method achieves far lower CTR across all typologies, indicating that it manages to consistently target topic-bias pairs which have lower interaction probabilities and thus are less likely to be in a user's history. 

Notably, \textit{devout and diverse} users have the lowest CTR scores for both methods (0.42 and 0.33), outside of the politically disengaged \textit{bystanders}. As previously mentioned, this class is heterogeneous in its biases. For the NN method, the significantly lower CTR score shows that it fails to accurately model this distinct partisanship.

We recall that given the disparate intentions of the NN method and our \textit{FNPC} approach, the former should achieve low WD scores and the latter should achieve relatively higher scores.
\textit{Devout and diverse} users have the highest 1-Wasserstein Distance (WD) among political typologies for the NN approach (1.42), indicating again that the NN method fails to cater to their heterogeneous partisanship over various topics.
The NN method has an average WD of 1. This means that if a user's previous reading history on a given topic contains only articles with an absolute bias score of 1 or higher (any non-central partisan lean), the NN method on average will not recommend articles with an opposing perspective. 
Meanwhile for \textit{FNPC}, all political typologies observe WD scores of at least 1.5, meaning that recommendations are consistently greater than 1 point off. The standard deviation of WD for \textit{FNPC} is 0.65, indicating that \textit{FNPC} almost always recommends articles that are shifted by at least a single bias score ($1.64 - 0.65 \approx 1$) and regularly those that are shifted by more than two ($1.64 + 0.65 \approx 2.3$). A WD score of 2 indicates that a given history over a topic with a single prior article of bias -2, -1, or 0 will result in a recommendation with bias scores 0, 1, or, 2. We see that for the three most polar typologies (\textit{core conservatives}, \textit{country first conservatives}, and \textit{solid liberals}) \textit{FNPC} observes over two times higher WD scores than NN. It is worth noting that \textit{solid liberals} are exceptionally different from the two aforementioned conservative classes, and \textit{FNPC} still manages to accurately target independently diverse articles for both classes. This means that \textit{FNPC} has the capacity to target left-leaning users as the furthest neighbors of right-leaning users, and vice-versa.  

The fourth highest WD score for \textit{FNPC} belongs to \textit{devout and diverse} users (1.65). Unlike the three more polarized typologies, this class of users are some of the most heterogeneous in their partisanship. This implies that while serving pointedly diverse articles to more polar users is easier to achieve, \textit{CPC}-based methods can also model heterogeneous users and still achieve \textit{pointed diversity}: it does not have an inverse relationship between WD trends and the heterogeneity of the associated typology, as the NN method does.

The average \textit{political tolerance} (PT) of the typologies, as seen in Table \ref{tbl:stats}, shows how broad each typology truly is in their ideology across topics. While this distribution cannot be precisely matched in scope by the RSs over limited recommendations, it can act as a target for the normalized entropy (NE) of our recommendations. \textit{FNPC} exhibits the highest normalized recommendation entropy (0.70), and thus most closely matches the spread of the true political tolerance (PT) of users. Unlike the NN method, the entropy it achieves is consistent across all typologies.

The two methods that target \textit{opposition users}, FN-NN and our method \textit{FNPC}, both exhibit global AB closer to 0. They also tend to be slightly left-leaning on average across most typologies. This is likely due to the fact that \textit{solid liberals} are the majority class (20\%) and tend to have consistently left-leaning biases across topics \cite{pew_typology}. As such, they may be more distinct from most users and more readily targeted as furthest neighbors. It is worth noting that \textit{solid liberals} are the only typology with right-leaning AB scores (0.05), indicating that our \textit{CPC} embedding space is able to identify that they are `more left' than the rest of the user space. 

\begin{figure}[htp]
    \centering
    \includegraphics[width=\columnwidth]{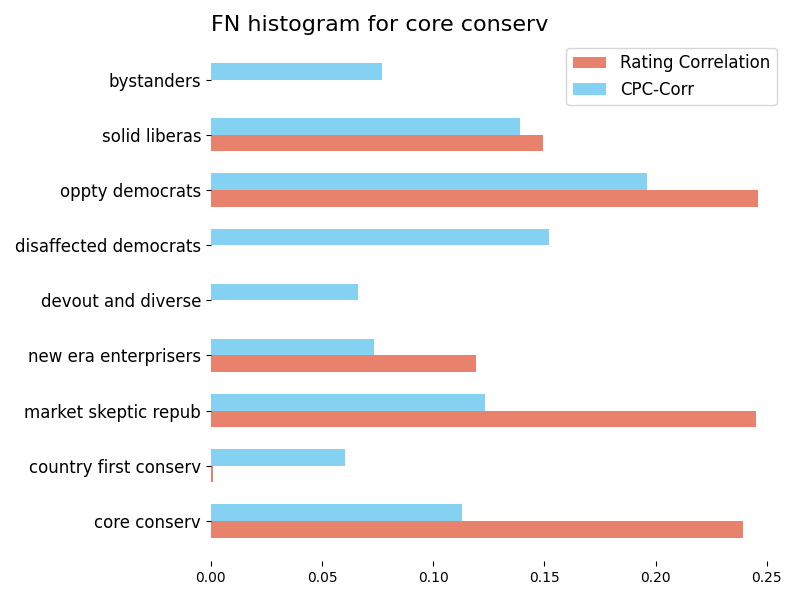}
    \caption{The average FN distribution of core conservatives induced by \textit{CPC}-based correlation and rating correlation.}
    \label{fig:fn_dist}
\end{figure}

Due to the fact that FN-NN has a global AB closer to 0, one might be tempted to say it achieves better \textit{pointed diversity}, but further investigation shows that this is not the case. 
Figure \ref{fig:fn_dist} shows the distribution of users within each typology that were chosen as furthest-neighbors on average by the core conservative class. The rating correlation used in FN-NN (shown in red) completely ignores four of the classes, and is highly focused on \textit{opportunity democrats}, \textit{market skeptic republicans} and \textit{core conservatives} (despite the class in question being \textit{core conservatives}, almost a quarter of the users targeted are of the same typology). 

Table \ref{tbl:extreme_FN_NN} shows an example of the adverse effects of this. The original interaction history over all \textit{core conservatives} recommended 31 articles on the topic of technology with an average bias of 0.33, and a relatively wide standard deviation. The NN method characteristically pushed this average bias further to the right and under-recommended the topic, as the topic of technology is approximately 2\% of the overall item distribution. \textit{FNPC} correctly targeted a left-leaning average bias close to -1, maintaining a similar standard deviation as the historical interactions. FN-NN recommended a much larger 107 articles on this topic, with every single one of them being of score -2. We believe that this is due to the topic of technology being of particular interest to \textit{market skeptic republicans}, who are notably left on this topic and evidently over-targeted as FNs via correlation over ratings. This example underlines the high polarity-seeking nature of the FN-NN approach, and the poor identification of furthest-neighbors according to a user's actual topical interests. 
It is worth noting that the furthest-neighbor targeting distribution induced by the rating matrix was effectively static for every single typology, underlining the ineffectuality of ratings alone to capture true political correlations.

\begin{table}
\centering
\caption{Example of the extreme polarizing nature of FN-NN on the topic of technology for core conservatives.}        \label{tbl:extreme_FN_NN}
\begin{tabular}{|l|c|c|c|c|}
\hline
 Stat & History & NN & FN-NN & FNPC\\
 \hline
 Avg. Bias &  0.33 $\pm$ 1.05 & 0.79 $\pm$ 1.16 & -2.00 $\pm$ 0 & -0.91 $\pm$ 1.01\\
 \hline
 Num. Recs. & 31 & 15 & 107 & 46\\
 \hline
\end{tabular}
\end{table}

Figure \ref{fig:fn_dist} also shows the average FN targeting of the core conservatives when using \textit{CPC} to calculate user correlation (blue). We see that these neighbors are sourced from throughout the political space and focused on left-leaning typologies, which is expected for the core conservative class. The targeting of every typology is characteristic of a method which is focused on modeling bias specifically over the relevant topics in a user's history, and not just the set of topics that are popular with the dominant user-base, as in the rating-based methods.

The high NE and WD values achieved by \textit{FNPC} across \textit{all typologies} indicates that the method targets articles that are distinct in their bias compared to a user's prior reading (high WD), are both moderate and polarized in their leanings (high NE), and that it achieves this for polarized users as well as those with heterogeneous partisanship. We conclude that the \textit{pointed diversity} achieved by \textit{FNPC} is due to the informativity of our \textit{CPC}-space.

\section{Discussion}

It is important to note that the low NE values achieved by the NN method, as shown in Table \ref{tbl:stats}, are indicative of recommendation behavior that would perpetuate filter bubble phenomena: given user histories generated from very broad \textit{political tolerances} (0.92 entropy on average), this method produces notably low-entropy recommendations (0.50 average), especially for the majority typologies (NN recommendations to \textit{solid liberals} have a normalized entropy of 0.36). Subsequent RSs trained on histories induced by this NN method would then only further perpetuate this ideological tightening. Meanwhile, \textit{FNPC} was provided identical ratings and histories, and maintained far greater entropy in recommendations. We argue that the \textit{pointed diversity} achieved by \textit{CPC}-based methods could help interrupt the polarizing positive feedback loop induced by iterative NRS deployments. 

Achieving state-of-the-art performance in managing the accuracy-diversity trade-off was not the motivation behind our contributions. Rather, we purposefully used simple, low-level approaches to isolate for the effects of \textit{CPC}-space. Our goal was to show that by directly modeling the biases of users relative to the content they enjoy, we can, in turn, define a space in which directional deviation correlates to a difference in perspective. We then showed that this method can be used to combat and counteract the perpetuation of prior biases that have been identified within NRSs. Future iterations of this system would benefit from using the depolarized embeddings in downstream tasks to filter for the most applicable topics.

The \textit{CPC} embedding space extends the default collaborative filtering paradigm in many powerful ways. First, it is relatively robust to sparsity in ratings and histories, unlike rating-correlation approaches. So long as there is a small collection of interacted articles, a meaningful and rich polarized embedding can be generated for a user regardless of the size of the global item set. More generally, it does not necessarily require any set of users to have histories over any common articles in order to place them relative to one another within \textit{CPC}-space. This is a powerful trait that addresses a long-standing limitation of CF methods in news recommendation, an environment where articles go stale quickly and users only interact with a small number of them at a time. 

Although we could have used a version of \textit{CPC} over nearest-neighbors to target maximum CTR, we consider such a system to be unethical and the antithesis of what we are trying to achieve with this work. As such, we apply a vanilla NN method to characterize the bias exploiting nature of common RS paradigms, thus establishing the problem in order to explain the value of how \textit{FNPC} addresses it. To validate that merely targeting furthest-neighbors via rating correlations was not strictly enough to achieve our goal, we also considered some characteristics of FN-NN. Ultimately, the weaknesses of the CF methods in NRS mentioned above are already reasonable incentives for using \textit{CPC}-based correlation, given the results we showed regarding its informativity. 

\textit{CPC}-based methods need not be limited to CF approaches and could be used in any system that benefits from modeling users relative to their biases.
The method is general in that it requires no knowledge of article topics or user typologies, and leverages bias labels associated only to the news outlet itself, not the articles directly. As such, this method could be applied to any language or set of topics, as long as a notion of news outlet alignment is available. The models as we have applied them are small and have minimal user history or data requirements. 
Thus \textit{CPC} can be easily recomputed in a variety of news environments.

\section{Conclusion}

We have proposed a novel embedding space, \textit{Constructed Political Coordinates}, where directional correlation corresponds to ideological similarity.
Through experimentation, we have shown that \textit{CPC} captures relevant information to understand an individual's biases over various topics. When applying \textit{CPC} to extract recommendations from politically diverse users, the recommendations it generates maintain high entropy over different biases and target specific biases that are diverse from a user's prior reading over a given topic. We achieve this pointed diversity without overly recommending extreme biases and while respecting the interests of heterogeneous political typologies. 

In this paper, we chose simple baselines mainly to showcase the capability of our \textit{CPC}-space. In the future, we would like to compare with more state-of-the-art baselines. We intend to generalize the identification of our landmark users to an automatic approach using the properties of our axis-aligned bounding box discussed in Section \ref{CPC}. We also intend to explore more robust GNN architectures for interest modeling. 

\section*{Acknowledgment}

We thank the anonymous reviewers for their helpful feedback on this work. We also gratefully acknowledge the financial support of the Natural Sciences and Engineering Research Council of Canada (NSERC).

\balance

\bibliographystyle{ieeetr}
\bibliography{sample-base}

\clearpage
\onecolumn

\end{document}